# An Arbitrary Time Interval Generator Base on Vernier Clocks with 0.67 ps Adjustable Steps Implemented in FPGA

**J. Wu**[a,1]

[a] *Fermi National Accelerator Laboratory,*
*Batavia, IL 60510, USA.*

*E-mail*: jywu168@fnal.gov

ABSTRACT: In TDC testing or timing system implementation tasks, it is often desirable to generate signal pulses with fine adjustable time intervals. In delay cell-based schemes, the time adjustment steps are limited by the propagation delays of the cells, which are typically 15 to 20 picoseconds per step and are sensitive to temperature and operating voltage. In this document, a purely digital scheme based on two vernier clocks with small frequency difference generated using cascaded PLL is reported. The scheme is tested in two families of low-cost FPGA and 0.67 and 0.97 picoseconds adjustable steps of the time intervals are achieved.

---

[1] Corresponding author.

# Contents

## 1. Introduction

In time-to-digital converter (TDC) testing[1-4], timing interval test signals with adjustable steps much finer than the TDC bin are desired. For timing system implementation, test signals with sub-picoseconds adjustable steps are also needed for various performance study and verifications.

It is possible to generate the test signals using analog methods, but the generated time intervals are sensitive to the circuit noise, grounding and power supply stabilities. In digital approaches[5-8], delay cell-based schemes usually do not have sufficiently fine adjustment steps. In today's FPGA devices and mainstream ASIC fabrication processes, delay cell propagation delays are typically 15 to 20 picoseconds. In fact, many TDC devices with such bin widths are to be studied or calibrated using sub-picoseconds time interval test signals. The propagation delay of the delay cells are also sensitive to the temperature and power supply voltage and if they are to be compensated using method such as delay-lock-loop (DLL), the cells must be designed to be slowed down with much larger propagation delays, that cause the time interval adjustment steps become even larger.

In this document, a purely digital scheme using vernier clocks with small frequency difference generated using cascaded PLL is reported. The scheme is shown in **Figure 1**.

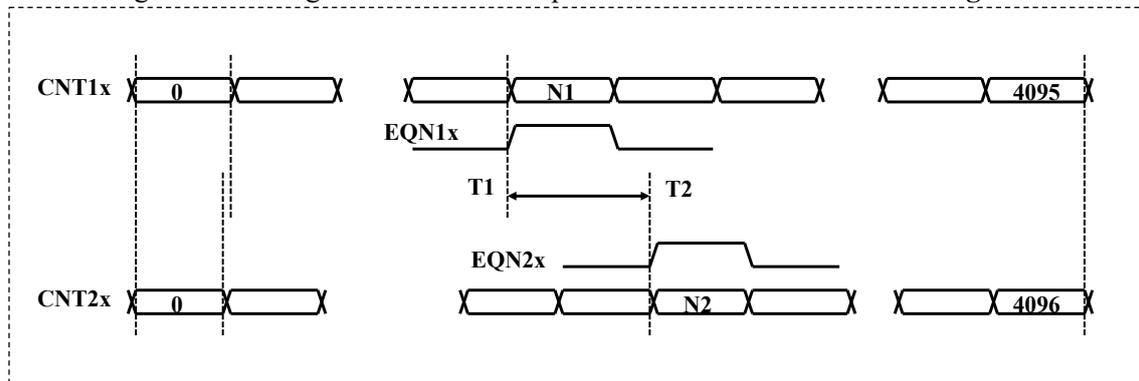

**Figure 1.** The dual clock time interval generation scheme

In the time interval generating circuit, two counters CNT1x and CNT2x, at least 12-bit each, are implemented and are driven by two clocks with small frequency difference. In the example shown above, the ratio of their frequencies is: $f_1/f_2 = 4095/4096$ and the two counters re-algins after 4095 periods of $f_1$. The output pulses are derived from two equality comparators that produce a pulse when CNT1x = N1 and CNT2x = N2. The time difference of the output pulse leading edges T2 - T1 can be adjusted by setting N1 and N2 with 0.97 ps or 0.76 ps steps when the clock frequency is 250 or 320 MHz in our tests. These two clocks are generated using internal cascaded phase-lock-loop (PLL) resources inside FPGA. The time interval generator is implemented and tested in an Altera/Intel Cyclone 5 [9] FPGA (5CGXFC5C6F27C7N) evaluation module (Terasic C5G [10]) and in the CMS ETROC emulator module with an Altera/Intel Cyclone 10 [11] FPGA device.

In Section 2, the implementation of the cascaded PLL is discussed and design details are described in Section 3. Test results are reported in Section 4 and conclusions are drawn in Section 5.

## 2. Implementation of the Cascaded PLL

The two clocks used in the time interval generator must be derived from a single oscillator source so that they are correlated with a known frequency ratio. There are several methods to generate clocks with fine adjustable frequencies but the author has only verified the method using integer PLL. Since the internal voltage controlled oscillator frequencies are limited within certain range, a single stage of the PLL will not be sufficient to produce clock signals with fine frequency differences. To produce frequency ratios as fine as 4095/4096 in our example, two PLL stages must be cascaded as shown in **Figure 3**.

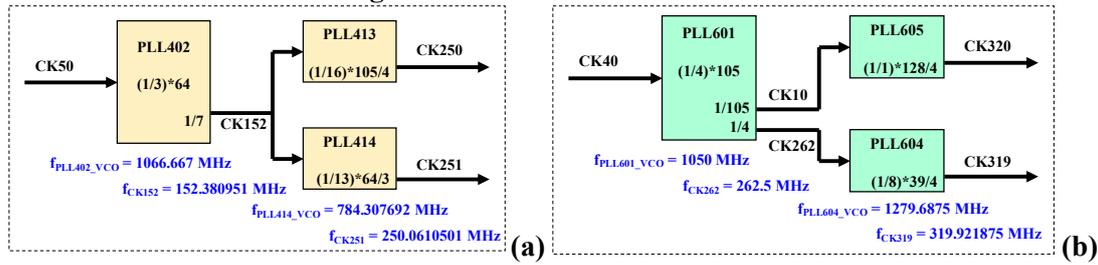

**Figure 2.** The cascade PLL structure: (a) in Cyclone 5, (b) in Cyclone 10 FPGA devices

Even with two PLL stages, multiplier integers must be carefully distributed among the PLL blocks to meet several constraints. One of the reasons of the ratio 4095/4096 is chosen is because 4095/4096 = (63*65)/(64*64) so that both the nominator and the denominator can be factorized and the factors can be distributed among the PLL blocks without violating design rules.

The cascaded PLL schemes are implemented in an Altera/Intel Cyclone 5 FPGA device and an Altera/Intel Cyclone 10 FPGA device (10CX220YF780E5G) with parameters shown above. In the Cyclone 5 implementation, external 50 MHz clock first produce in intermediate clock CK152 and then two clocks CK250 and CK251 with frequency 250 MHz and (4096/4095)*250 MHz, respectively, are produced. The clock period difference is approximately 0.976 ps. In the Cyclone 10 implementation, external clock frequency is 40 MHz and two clocks CK320 and CK319 with frequency 320 MHz and (4095/4096)*320 MHz, respectively are produced. The clock period difference is approximately 0.763 ps.

Modern FPGA design software may not allow the designers to enter the parameters of the PLL directly. So sometimes the designers may need to "cheat" the software to reach needed parameter. Some FPGA families may support fractional PLL mode which will be verified in future studies, but for now, the PLL is forced to be in integer mode.

## 3. Design Details

The time interval generator consist of two sets of rotational counters, constant registers, and equality comparators. The block diagram of the time interval generator is shown in **Figure 3**(a).

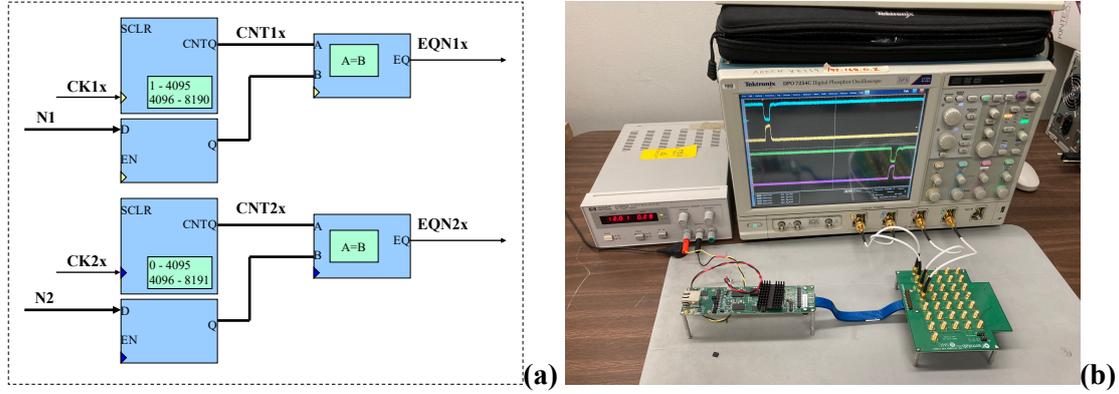

**Figure 3.** (a) Block diagram of the time interval generator (b) Test setup

Implementing 12-bit rotational counters should be sufficient but we actually implemented 13 bits for convenience in the tests. One of the counter CNT2x is a plain binary counter that runs from 0 to 8191 rotationally and is driven by the faster clock CK2x. Another counter CNT1x is driven by the slower clock CK1x so it counts from 1 to 8190 and repeats. During initialization, the counters are synchronously set to 4096 when CK1x and CK2x are approximately aligned.

The constant registers brings integers N1 and N2 from slow control clock domain to the CK1x and CK2x clock domains, respectively, so that the comparators in the later stage will always output clean pulses at the intended time.

The comparators compare the running counter values CNT1x and CNT2x with the registered N1 and N2 constants and output pulses when CNT1x = N1 or CNT2x = N2, respectively.

The test setup with re-purposed CMS ETROC emulator module is shown in **Figure 3**(b). The outputs use in differential 1.2-V SSTL (Stub Series Terminated Logic) I/O standard. The oscilloscope is Tektronix DPO 7254C Digital Phosphor Oscilloscope with 10 G samples/s each channel at interpolate time mode. The positive and negative outputs of the differential pair of the EQN1x and EQ2x signal are fed into CH1 to CH4 of the oscilloscope, respectively.

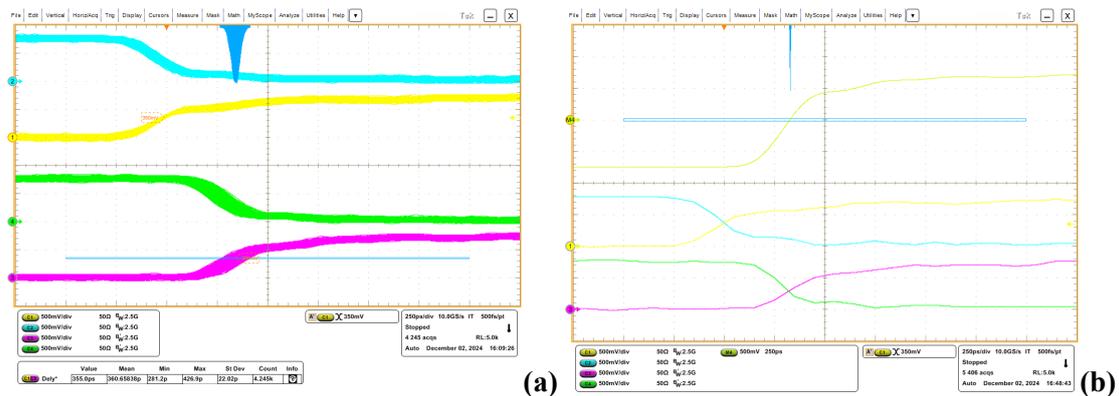

**Figure 4.** The outputs of the time interval generator: (a) two differential pairs and histogram of the time difference, (b) the average and its timing distribution histogram

The output of the FPGA PLL blocks have timing jitters around 22 ps as shown in **Figure 4**(a).

To see the time interval adjustment effects more clearly, the oscilloscope internal average function is used as shown in **Figure 4**(b), in which M4 = AVG(CH3 - CH4) where AVG is the function of making sliding averages over 400 samples. Clearly the timing distribution of the average is much narrower than the raw measurements.

The time interval is determined with the following equation:

$$T_2 - T_1 = \frac{N_2 - 4096}{f_2} - \frac{N_1 - 4096}{f_1} = \frac{N_2 - N_1}{f_2} - N_1\left(\frac{1}{f_1} - \frac{1}{f_2}\right)$$

When the two constants N1 and N2 increase or decrease by the same value, N2-N1 remains unchanged. In this case the time interval is fine adjusted with steps ($1/f_1 - 1/f_2$) as indicated by the second term in the equation above.

On the other hand, when N2 increases while N1 is unchanged, a "coarse" adjustment to the time interval is made as indicated by the first term in the equation. Large time intervals could be used in studies with long cable or optical fiber. With 13 bits, the difference between N2 and N1 can be about 8000, which will produce about 24 μs time intervals when $f_2$ = 320 MHz.

## 4. Test results

The tests are conducted using a simple sequencer implemented in the FPGA firmware with 16 pre-determined time interval values, repeating for about 429 seconds per point allowing the oscilloscope to make about 9600 acquisitions for each point. The sliding average mentioned earlier are shown in **Figure 5**(a) as the trace M4. Histograms of M4 leading edge time are booked and output as a file.

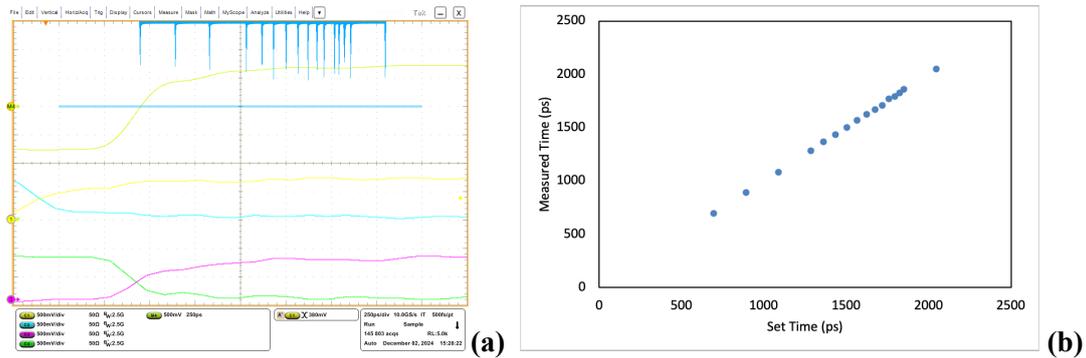

**Figure 5.** (a) Oscilloscope screenshot of the test, (b) Measured time vs. set time

The separations of the peaks are various, ranging from 24.2 to 195.3 ps, corresponding differences of 32 to 256 in constants N2 and N1 settings.

The center times of the peaks in the histogram are calculated using weighted average over 11 histogram bins around each peak. The measured time and the set time are compared in the plot shown in **Figure 5**(b).

The time offsets, i.e., the differences of the measured time and the set time are calculated and plotted in **Figure 6**(a).

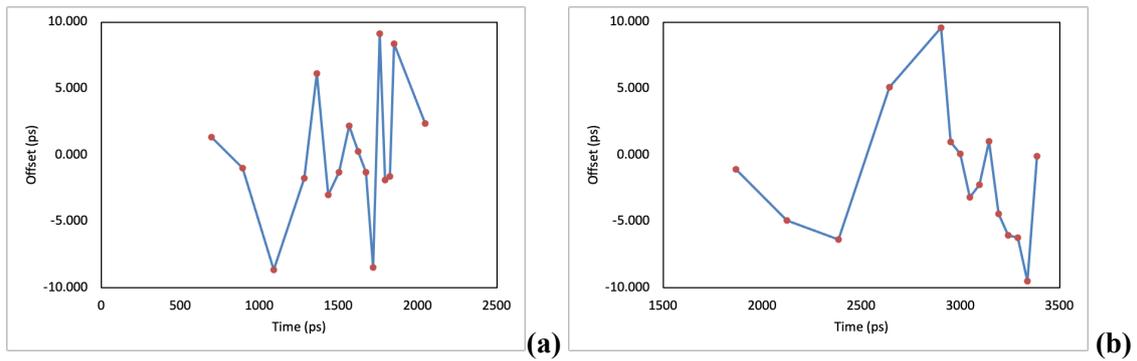

**Figure 6.** Time offsets of the test: (a) in the Cyclone 10 FPGA, (b) in the Cyclone 5 FPGA

The time offsets between the set time and measured time for our test in the Cyclone 10 FPGA within +- 10 ps are achieved. The offsets may come from interference among the PLL blocks via analog power supply coupling, since they share the same analog power rail inside the FPGA device. But this hypothesis must be studies in the future work.

The time interval generator implemented in the Cyclone 5 FPGA are also tested and the timing offset results are show in **Figure 6**(b). The time offsets withing +- 10 ps are also achieved.

The "coarse" adjustment mentioned earlier allows users to extend dynamic range indefinitely. In the test shown in **Figure 7**, N2-N1 was set to several selected values from 0 to 27 with $f_2$ = 320 MHz.

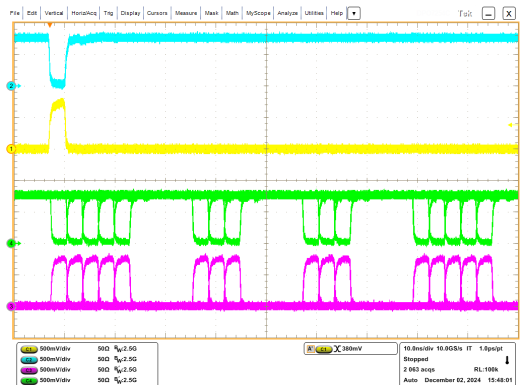

**Figure 7.** Time intervals over a large range

The oscilloscope was set in infinity persist mode and time intervals up to about 87 ns can be seen.

## 5. Conclusions

The feasibility of implementing time interval generator using two clocks with small frequency difference is verified. Adjustment steps of 0.97 and 0.76 ps are achieved.

The timing jitters and offsets are determined by the performance of PLL blocks and about 22 ps timing jitters and +- 10 ps timing offsets are achieved in both Cyclone 5 and Cyclone 10 FPGA families. It is expectable to further reduce the timing jitters and offsets using high quality external PLL products but it needs to be verified in the future work.

## Acknowledgments

This manuscript has been authored by Fermi Forward Discovery Group, LLC under Contract No. 89243024CSC000002 with the U.S. Department of Energy, Office of Science, Office of High Energy Physics.